# Transverse Beam Envelope Measurements and the Limitations of the 3-Screen Emittance Method for Space-Charge Dominated Beams.

J. G. Power[*], H. Wang, M. E. Conde, W. Gai, R. Konecny, W. Liu, and Z. Yusof

High Energy Physics Division, Argonne National Laboratory, Argonne, IL 60439, USA

Abstract

In its normal mode of operation the Argonne Wakefield Accelerator Facility uses a high charge (10-100 nC), short pulse (3-5 psec) drive bunch to excite high-gradient accelerating fields in various slow-wave structures. To generate this bunch, we designed a 1.5 cell, L-band, rf photocathode gun with an emittance compensating solenoid to give optimal performance at high-charge; it has recently completed commissioning. More recently, we have begun to investigate the possibility of using this gun in a high-brightness, low-charge operating mode, with charge equal to approximately 1 nC, for high-precision measurements of wakefields. Two related measurements are reported on in this paper: (1) measurements of the transverse beam envelope are compared to predictions from the beam dynamics code PARMELA; and (2) investigations into the use of a modified 3-screen emittance measurement method that uses a beam envelope model that includes both space-charge and emittance effects. Both measurements were made for the 1 nC, 8 MeV beam in the drift region directly following the rf photocathode gun.

---

[*] jp@anl.gov

## 1. Introduction

The Argonne Wakefield Accelerator (AWA) facility [1] is used to investigate electron beam-driven, high-gradient wakefield acceleration techniques. The generation of high accelerating gradients (>100 MeV/m) in wakefield structures or plasmas requires a high charge (10-100 nC), short pulse (3-5 psec) electron drive bunch. While most rf photocathode gun designs are optimized for high-brightness, low-charge (~1 nC) applications, such as free electron lasers, the 1.5 cell, L-band AWA rf photocathode gun, or "Drive Gun," is optimized for high-charge operation mode. The high-brightness application we are pursuing is the development of a compact, high-precision wakefield measurement system for the purpose of characterizing slow-wave structures as described in the reference [2].

Recently, we have begun an investigation to determine if the high-charge AWA Drive Gun can also be operated in a high-brightness operation mode. High-brightness rf photocathode guns routinely generate 1 nC beams with normalized transverse emittances in the neighborhood of 2 – 5 μm. Since the Drive Gun had already been optimized for high-charge operation, we realized that it might not be possible to generate such high-brightness beam with this gun. However, recent simulations [2] performed with PARMELA [3] indicate that high-brightness operation is possible provided that a sufficiently high accelerating gradient (peak field on-axis = $E_{z0} = 80 MV/m$) in the gun can be reached.

In this paper, we present the results of experiments to characterize the AWA Drive Gun operating in the high-brightness, low-charge mode are presented. Spot size measurements were made of the 8 MeV, 1 nC beam at 3 locations in the drift region downstream of the Drive Gun as the rf launch phase of the gun was varied from approximately $30^0$ to $70^0$. These measurements were then analyzed in two different ways: (1) the spot sizes were compared to predictions from the beam dynamics code PARMELA; and (2) the spot sizes were used to investigate the capabilities of a modified 3-screen emittance measurement method. This modified method differs from the standard method in that its beam envelope model includes space-charge effects in addition to emittance effects, since it is well known that standard method fails for space-charge dominated beams. The outline for the remainder of this paper is as follows. We present: (i) the experimental setup and technique used to measure the beam envelope and emittance, (ii) the results of the experimental measurements; and (iii) a discussion of results.

**2. The Experimental Setup and Technique**

In this section, we first describe the experimental setup including the main components of the beamline where the spot size measurements were made, the Drive Gun, and the photocathode laser system; and second the experimental technique of the modified 3-screen emittance measurement method.

*2.1 The AWA Gun Test Stand*

In Figure 1, a highly simplified version of the AWA Gun Test Stand (GTS) beamline shows only the elements that were used during these measurements. The block labeled "Gun & Solenoids" consist of the Drive Gun and three solenoids as described in detail below. The two diagnostics elements used are: (1) a single Bergoz integrating current transformer (ICT) for nondestructive measurement of the beam charge; and (2) three YAG:Ce (YAG) phosphor screens for beam profile measurements (YAG-1, YAG-2, and YAG-3). The YAG screens were mounted on pneumatic actuators, were purchased from Crytur Ltd., and have dimensions of 250 $\mu$m thick by 50 mm in diameter.

THE RF PHOTOCATHODE GUN

The AWA Drive Gun [1] is similar in design to the standard BNL/SLAC/UCLA rf photocathode gun with an emittance compensating solenoid. The Drive Gun is an L-band, 1.5 cell, rf photocathode gun operating in the $\pi$-mode at 1300 MHz and currently uses a Mg photocathode. Power from a 30 MW L-band Klystron is side-coupled into the downstream full cell and then on-axis coupled into the upstream half cell. Two solenoids, the bucking and the focusing solenoids, are placed symmetrically about the plane of the photocathode while a third solenoid, the matching solenoid, is located at the exit of the gun. The focusing and matching solenoids work together as the effective emittance compensating solenoid, while the bucking solenoid is used to buck out the magnetic field of the focusing solenoid at the cathode. We have recently implemented a laser cleaning system and initial estimates show that the quantum efficiency was improved by about an order of magnitude.

The Drive Gun has been successfully conditioned and routinely generates electron bunches with charge up to 100 nC. Using 13 MW of incident power, the gun is normally operated at $E_{z0} = 80 MV/m$ which generates a beam energy at the gun exit of about 8 MeV. The bucking and focusing solenoids are operated with on-axis peak magnetic flux density of $B_z = 0.92 kG$ while the matching solenoid runs at $B_z = 4.5 kG$. Under these conditions, the emittance compensated minimum is located approximately 2.5 m downstream of the photocathode.

THE AWA PHOTOCATHODE GUN LASER SYSTEM

The AWA laser system consists of a Tsunami oscillator purchased from Spectra Physics and a TSA-50 chirped-pulse amplifier (CPA) from Positive Light. The oscillator is a regenerative mode-locking Ti:Sapphire oscillator tuned to λ = 744 nm (Δλ = 9 nm, E = 7 nJ) output that is pumped by an all solid-state 5 W, cw, green laser (Millennia V). The oscillator runs at the 16[th] sub-harmonic (81.25 MHz) of the rf (1300 MHz) and is phase locked to within 1 psec (r.m.s. jitter) of the rf using the vendor's *Lok-to-Clock* box. The CPA consists of a stretcher followed by a Ti:Sapphire regenerative amplifier (regen), two stages of linear Ti:sapphire amplifiers and a compressor. After reducing the bandwidth with a mask in the stretcher, the seed pulse exits the stretcher with a pulse length of approximately 200 psec where it is optically switched into the regen with the injection Pockels Cell, at a repetition rate of 10 Hz. The extraction Pockels Cell timing is adjusted to extract the most stable pulse from the regen and this pulse (E = 2.5 mJ) is then further amplified by the two linear amplifiers to 35 mJ and directed into the compressor. Upon exiting the compressor the pulse energy is 20 mJ and the pulse length is 8 psec FWHM. After exiting the CPA, this pulse is frequency tripled to $\lambda = 248$ nm and $E = 1 - 2$ mJ, in the Tripler.

Upon exiting the Tripler, the laser pulse passes through a variable telescope and is then transported by a series of mirrors into the accelerator tunnel to an adjustable iris, about 10 m after the Tripler. The adjustable iris is located just outside the vacuum, about 1 m away from the photocathode surface, and is used to control the spot size on the photocathode. After the nearly parallel beam passes through the iris, it is sent through a UV-quality quartz window and strikes a mirror in the vacuum chamber and is directed at near normal incidence to the photocathode.

THE MODIFIED 3-SCREEN EMITTANCE MEASUREMENT METHOD

The standard 3-screen emittance measurement method is not adequate for measuring the emittance of the beam directly out of the Drive Gun due to the strong space-charge forces involved in the transport of the beam. In the standard method, space-charge is implicitly ignored and one determines the three transverse phase space unknowns (the beam emittance ($\varepsilon$), the beam beta function ($\beta$), and the beam envelope slope ($\alpha$)) from the 3 measured spot sizes ($\sigma_1$, $\sigma_2$, and $\sigma_3$). However, when the space-charge cannot be ignored, a modification to the standard method will be required.

A criteria to determine when it is appropriate to neglect space-charge can be obtain from the 1D beam (infinitely long, round beam) RMS envelope equation [4],

$$\sigma''(s) + \kappa_0(s)\sigma(s) - \frac{\varepsilon^2}{\sigma(s)^3} - \frac{K}{4\sigma(s)} = 0 \quad (1)$$

where s is the direction of propagation, the prime means the derivative is taken with respect to s, $\sigma$ is the RMS envelope or radius of the beam, $\kappa_0(s)$ is the external focusing term, $\varepsilon$ is the unnormalized transverse emittance, and $K$ is the space-charge parameter (or generalized perveance) which is given by $K = 2I/(I_0\beta^3\gamma^3)$ where $I$ is the peak current, $I_0$ = *17000* Amps is the electron characteristic current, and β and γ are the usual relativistic factors. The third term in Eq. 1 is called the emittance defocusing term while the fourth term is called the space-charge defocusing term. If all three measurements of the beam size are made in a drift (as is the case for the measurements described in this paper) we can set $\kappa_0(s)=0$ and the evolution of the beam envelope is determined solely by the emittance and space-charge terms. If we define *R* as the ratio of the space-charge to the emittance terms we then have,

$$R = \frac{K\sigma^2}{4\varepsilon^2} \quad (2)$$

The standard 3-screen emittance measurement technique is used when the beam is emittance dominated; i.e. when this ratio is much less than one: $R \ll 1$.

According to PARMELA simulations of the AWA Drive Gun, a $Q = 1$ nC, Energy = 8 MeV beam, will have: $\sigma_{x,y} \sim 1\text{-}3$ mm, $\varepsilon_{norm} \sim 1$ µm, and $I \sim 0.5$ kAmp, which results in $R \sim 20\text{-}150$, clearly a space-charge dominated beam. Therefore, to extract the emittance from the spot size measurements we need to use a model of the beam envelope that includes both space-charge and emittance; there are several ways to do this. In this paper, we will fit the beam envelope to the 3 spot size measurements with the 3D envelope code TRACE 3-D [5]. Since there are 4 unknowns in Eq. 1 (the initial spot size, $\sigma(s=0)$, and slope, $\sigma'(s=0)$, the emittance ε and generalized perveance K) one could imagine using a 4-screen emittance measurement technique to determine the 4 unknowns. However, as we discuss below, we chose to assume that K is given correctly from PARMELA simulations.

### 3. Experimental Measurements and Emittance Estimates

*3.1 Data Acquisition and Data Reduction*

Beam images were acquired at 3 locations in the drift region following the gun, (Fig. 1: YAG-1, YAG-2, and YAG-3) with *Sony* analog CCD cameras fitted with manually controlled 60-300 mm zoom lenses and captured by an 8-bit PC-based *Imaging Technology* frame grabber. The zoom lenses were instrumented with a remotely controllable aperture to insure that the images were not saturated and yet made full use of the dynamic range of the imaging system. A user developed Matlab routine was used to extract spot sizes ($\sigma_x$ and $\sigma_y$) from each of the captured beam images using the following procedure: (1) subtract off the dark current background from the image, and (2) fitting a Gaussian curve to the projected beam profile. All spot sizes quoted in this paper are the '1-sigma' values of the fitted Gaussian.

The spatial resolution of the acquired images was estimated as follows. The horizontal resolution of the image was 11.6 pixels/mm while the vertical resolution was about 8.1 pixels/mm. (The difference in resolution for the two directions is due to the fact that the YAG screens are tilted at $45^0$ which caused the horizontal axis to become elongated by $\sqrt{2}$.) In units of distance, and assuming we need a least 2 pixels to detect a change in spot size, then the horizontal resolution is = (2 pixels)/(11.6 pixels/mm) = 0.09 mm and the vertical resolution is 1/8.1 = 0.12 mm. (For simplicity of data analysis we use 0.1 mm in both directions.) Since we need at least 10 pixels per 1-sigma of a Gaussian beam to resolve the spot [6], this means that the minimum 1-sigma spot size that can be resolved in the horizontal is 10/11.6 = 0.86 mm and in the vertical direction is 10/8.6 = 1.2 mm.

During the experiment, we attempted to hold all experimental parameters fixed except for the rf gun phase, $\phi_{gun}$, as shown in Table 1. However, due to the inherent shot-to-shot jitter of the laser energy amplitude, the nominal charge, $Q = 1\,nC$, was expected to fluctuate during the data taking process. And since charge jitter causes the spot size to jitter, data analysis becomes difficult. On the other hand, we expect that beam charge and spot size to be closely correlated, assuming the other parameters remain fixed. Therefore, rather than acquiring a large number of images and averaging the results, we implemented a method for synchronized capturing of the beam image (as measured at the phosphor screens in Fig. 1: YAG-1, YAG-2, and YAG-3), and the value of charge (Fig. 1.: ICT). Since the other parameters can be held constant rather easily, the synchronized data taking procedure was expected to greatly reduce the measured spot size fluctuation due to laser amplitude jitter.

After setting the parameters to the values shown in Table 1, data was acquired by the following procedure. (i) Make a quick sweep of the rf gun phase to determine the well understood charge vs. rf gun phase (Q vs. $\phi_{gun}$) curve. From this curve, the approximate location of the zero phase point, $0^0$, is found since it is the approximately the lowest phase point at which the extracted charge from the gun drops to zero. Once the zero phase point is found, increase $\phi_{gun}$ to approximately $30^0$; (ii) Insert YAG-1 into the beamline; (iii) Over approximately the next 20 shots of the machine, acquire the video image at YAG-1, the charge at ICT, and the relative intensity of the laser pulse (measured at the exit of the laser Tripler); (iv) Repeat steps (ii) and (iii) for the other phosphor screens (YAG-2 and YAG-3) with the same rf gun phase; (v) increase $\phi_{gun}$ by $10^0$ and repeat steps (i) – (iv). In summary, we measured the spot sizes at YAG-1, YAG-2, and YAG-3 for the rf gun phases $\phi_{EXPT} \cong 30^0, 40^0, 50^0, 60^0$, and $70^0$. It should also be emphasized that the solenoids used for emittance compensation were not adjusted during this entire run which means that the optimal (i.e. very low) emittances were not obtained since the beam is not, in general, emittance compensated.

A typical data set (20 machine pulses acquired) of the spot size vs. charge is shown in Figure 2. Although our goal was to measure the spot size at Q = 1 nC, in general, there were no events at precisely Q = 1 nC and some fitting is required. Due to laser amplitude jitter as discussed above, we observed a large variation in the beam charge over the entire range of data, from $Q_{min}$ = 0.8 nC to $Q_{max}$ = 1.6 nC, as well a large variation in the 1-sigma Gaussian fitted spot size measured at YAG-1, from $\sigma_{min}$ = 1.7 mm to $\sigma_{max}$ = 2.7 mm. Taken as they are, these large variations make data interpretation difficult. However, this variation can be reduced by using only spot sizes in a narrow charge window surrounding Q = 1 nC (Fig. 2a. 0.95 nC < Q < 1.05 nC). For this window, we obtained a significant reduction in the peak-to-peak spot size fluctuations, $\Delta\sigma_{p-p}$ = 0.4 mm. However, note that it is also possible to choose a narrow charge window that still contains large spot size fluctuations (Fig. 2b). So, although the windowed data usually gave a marked improvement over the non-windowed data, the remaining variation in the spot size is not fully understood at present but investigations are underway.

In addition to the complication of spot size fluctuations, there were usually not enough data points in the window (e.g. there are only 3 data points in the charge window shown in Fig. 2a) to reliably obtain a mean. Therefore, a linear fit to the data was made (solid line in Fig. 2) and the value for the spot size was interpolated at Q = 1 nC. In the case of Fig. 2, the spot size obtained at Q = 1 nC was $\sigma_x$ = 1.97 mm. The measured horizontal and vertical spot sizes (using the above procedure for the data reduction) for the 3 different YAG screens and the 5 different phases are shown in Figure 3.

*3.2 Comparison of the Measured Spot Sizes to the PARMELA Beam Envelope.*

In this section, we compare the measured spot sizes to PARMELA predictions of the beam size at the 3 YAG screens shown in Figure 1 using the machine and beam parameters listed in Table 1. Since the precise value of the rf gun phase is not known during the experiment, the approximate values of the rf gun phases listed section 3.1 are replaced with fitted values in this section. The fit was done by adjusting the rf gun phase in the PARMELA simulation until the envelope most closely matched the $\phi_{EXPT} \cong 30^0$ data points. The fit resulted in a small adjustment of $2^0$ of rf phase so that the set of phases quoted above ($30^0$ to $70^0$) are now replaced with the set ($28^0$ to $68^0$).

Measured spot sizes and the PARMELA beam envelope are in reasonably good agreement in the drift region after the gun. For instance, by inspection of Figure 6a, one can see that the PARMELA beam envelope for $\phi_{PARM} = 28^0$ passes very closely to the three $\sigma X$ data points for $\phi_{EXPT} = 28^0$ (square symbols). However, a careful inspection of data in Figure 6 reveals two systematic anomalies with the spots measured at YAG-1 (z = 102 cm). First, the ratio of $\sigma Y$ to $\sigma X$ at the three screen locations have the average values of $\sigma Y/\sigma X$ = 1.16, 1.02, and 1.00, respectively. This means that while the spots measured at YAG-2 and YAG-3 are essentially round, the spot measured at YAG-1 is elliptical. The second anomaly is that the spot size measured at YAG-1 is systematically larger than the predicted PARMELA envelope in both the horizontal and vertical plane, while it is within a few percent at YAG-2 and YAG-3. On average, the measured value of $\sigma X$ at YAG-1 is 16% higher than the predicted PARMELA beam envelope while the measured value for σY is 34% higher. Both of the systematic errors just described are believed to be caused by the large dark-current background present at YAG-1 due to its close proximity to the Drive Gun. On the contrary, images captured at YAG-2 and YAG-3 have almost zero dark-current background, since they are located farther downstream, and thus it is easier to obtain an accurate spot size measurement. In the future, the problem due to dark-current encountered at *YAG-1* could be abated by using an OTR (optical transition radiation) screen and a gated camera to effectively gate out the dark current.

*3.3 TRACE 3-D fit to the Data*

In this section, we use the 3D beam envelope code, TRACE 3-D, to fit a beam envelope to the spot sizes for the 5 different rf gun phases (Fig. 3) in order estimate the emittance. In the fit, the transverse phase space parameters in the horizontal ($\alpha_x$, $\beta_x$, and $\varepsilon_x$) and vertical ($\alpha_y$, $\beta_y$, and $\varepsilon_y$) directions are varied to obtain a best fit of the envelope to the data for each plane. The longitudinal parameters (bunch length, energy, and energy spread) are taken from the PARMELA simulations and used as inputs to TRACE 3-D. This is acceptable for two reasons: (i) previous measurements of the longitudinal parameters have been found to be in reasonable agreement with PARMELA predictions; and (ii) the transverse parameters are only weakly dependent on the longitudinal parameters in our range of interest. An additional simplification used in the fit is that TRACE 3-D only tracks the overall beam envelope, and not individual longitudinal slices, and therefore doesn't correctly model the evolution of the projected emittance.

Lastly, the longitudinal parameters are approximated to be constant and equal to the averaged values of the PARMELA output in the drift region between YAG-1 at z = 102 cm and YAG-3 at z = 375 cm. Since the simulations show that the longitudinal parameters only varied by a few percent in this interval a good fit is expected.

The beam envelope is fitted to the data in the following manner. If $\sigma_i$ are the measured spot sizes and $T_i$ are the TRACE 3-D fitted spot sizes at the three YAG screens, then $\chi^2$ is defined as the sum of the square of the residuals, $\chi^2 = \sum_{i=1}^{3}(\sigma_i - T_i)^2$. Normally, a best fit to the data is obtained by varying the three phase space unknowns {ε, β, α} in TRACE 3-D until $\chi^2$ is minimized, and this is essentially our approach. However, it is more meaningful to find two sets of parameters (ε, β, α) that both pass within the error bars of the data in order to obtain a lower and an upper estimate of emittance.

As a representative example of the quality of the TRACE 3-D fit to the data, we examine the fit to the horizontal (x) data for $\phi_{gun} = 28^0$ in some detail (Fig. 5). The error bars on the spot size measurements come from two factors, the resolution of the video system (which was shown to be +/- 0.1 mm in Section 3.1) and the spot size uncertainty due to shot-to-shot fluctuations of the spot size (i.e. spot size jitter) as shown in Figure 2b. This later fluctuation is difficult to estimate given the limited number of data points, but the typical standard deviation of the spot sizes over the full charge range was approximately 0.3 mm. Therefore, adding these errors in quadruture gives the total error bars as approximately +/-0.3 mm (Fig 5: note that the origin has been shifted so that YAG-1 is located a z = 0 cm.) As can be seen from the figure, there is a large emittance range that fits the data for the case at $\phi_{gun} = 28^0$. These fit results mean that the normalized emittance is bounded between 1 μm and 13 μm for this case.

The range of the normalized r.m.s. emittance for all 5 values of the rf phase and for both the horizontal and vertical planes is shown in Figure 6. Although the fit is technically consistent with the PARMELA prediction, its emittance resolution was poor; i.e. the quality of the emittance fit is poor since it only estimates the emittance to about an order of magnitude. A full explanation of why the emittance was not more tightly bounded is given in the next section, but is basically due to the fact that the beam is emittance dominated, i.e. a large value of $R$ (Eq. 2) in the drift region.

## 4. Discussion of Results

Overall, agreement between the model predictions and the measured values were of mixed quality. On the positive, the measured beam spot sizes were in good agreement with the predicted values from PARMELA, except at YAG-1 where the dark-current background limited the resolution of the measurement. However, the modified 3-screen emittance measurement method used in this paper proved unsatisfactory since its emittance resolution was poor. This result seems surprising at first since the envelope equation upon which TRACE 3-D is based includes both space-charge and emittance effects and was therefore expected to be useful in constraining the beam emittance. However, the reason that the 3-screen technique failed was due to a combination of the large space-charge dominated ratio of the beam, $R$, and the accuracy with which we can measure the spot sizes at YAG-1, YAG-2, and YAG-3. A sensitive measurement of the emittance should do two things: (i) produce a large change in the measured variable (spot size in the 3-screen case) for a small change in emittance; and (ii) be able to measure the variable (spot size) with high precision. But for the 3-screen measurement, in the case of a heavily space-charge dominated beam, even large changes in emittance (e.g. a factor of 2) did little to change the beam envelope (or spot size), as can be seen from Figure 5. For example, when $R = 300$, as is typical for the beams measured in these experiments, then the space-charge force is 300 times stronger than the emittance force. In other words, the evolution of the beam envelope is almost totally uninfluenced by the low value of the emittance. According to Eq. 2, the emittance would need to increase by a factor of $\sqrt{300}$ (or ~17) to reduce $R$ to near 1 and emittance effects could be easily measured.

With the above explanation of the importance of $R$ in mind, we can now understand the variation of the emittance fit vs. phase shown in Figure 6. Inspection of Figure 6 reveals that the normalized emittance is more tightly bound at the lowest ($28^0$) and highest ($68^0$) values of the rf gun phase compared to the rf phase near the center ($48^0$). This is due to the variation of the space-charge ratio, $R$, for each of the different rf phases, as is shown for two representative examples in Figure 7. At $\phi_{gun} = 28^0$, the space-charge ratio $R$ has an average value of about 250 in the interval between YAG-1 and YAG-3, while at $\phi_{gun} = 48^0$ this average value of R is about 1000. From the above discussion, we know that the more heavily space-charge dominated a beam is, the less precision with which we can constrain the emittance. Therefore, since the space-charge ratio $R$ is larger at $\phi_{gun} = 48^0$ than at $\phi_{gun} = 28^0$ we expect that the 3-screen measurement to be less precise at $\phi_{gun} = 48^0$ and this is exactly what was observed (Fig. 6).

## 5. Conclusion

Two related measurements are reported on in this paper: (1) are compared to predictions from the beam dynamics code PARMELA; and (2) that uses a beam envelope model that includes both space-charge and emittance effects. Both measurements were made for the 1 nC, 8 MeV beam in the drift region directly following the rf photocathode gun.

    Beam dynamics studies were made for the space-charge dominated beam produced by the 1.5 cell, rf photocathode gun at the Argonne Wakefield Accelerator facility. Measurements of the transverse beam envelope were made in the drift region located between 102 cm and 375 cm after the photocathode plane, for a beam with charge of 1 nC, energy of 8 MeV, and a bunch length of 0.8 mm. The measurements of the beam envelope in the drift were in reasonably good agreement with the PARMELA predictions. Investigations into the use of a modified 3-screen emittance measurement method using a model that includes both space-charge and emittance effects were made. However, because the beam was so heavily space-charge dominated (R>200) this technique did not have sufficient emittance resolution to constrain the emittance to within a useful range. However, the modified 3-screen technique is still useful for measuring the emittance of a beam that is not so space-charge dominated, say when $R < 10$. In the future, we plan to measure the emittance of this beam using both a standard pepper pot technique and an ODR-OTR based technique [7].

## 6. Acknowledgments

This work was supported by the Division of High Energy Physics, US Department of Energy under contract No. W-31-109-ENG-38.

FIGURE CAPTIONS

Fig. 1. A simplified diagram of the AWA Gun Test Stand beamline. Charge is measured by the ICT and beam images are measured by inserting the phosphor screens YAG-1, YAG-2, or YAG-3. ($z = 0$ is the location of the photocathode; all distances are in cm)

Fig. 2. Spot size vs. charge measurements for a typical data event. (a) small fluctuations in spot size are obtained for the charge window between Q = 0.95 and 1.05 nC while (b) large fluctuations in the spot size for a different charge window between Q = 1.24 and 1.34 nC

Fig. 3. Spot size measurements made at the three YAG screens for Q = 1 nC for the 5 different rf gun phases used during the experiment ($\phi_{EXPT}$). (a) $\sigma X$ and (b) $\sigma Y$ are the 1-sigma fitted Gaussian spot sizes in the horizontal and vertical directions, respectively. Note, the lines are only used as a visual guide to connect spots measured at the same phase and are not meant to be representative of the beam envelope. Also note that the solenoids were set to a constant value (Table 1) during the rf phase sweep and thus the emittances are not, in general, compensated.

Fig. 4. Comparison of the PARMELA beam envelope (solid lines) and the measured spot sizes (symbols) shows relatively good agreement.

Fig. 5. Typical fit of the TRACE 3-D envelope (solid lines) to the $\sigma X$ spot size data (square symbols with error bars) for the particular value of $\phi_{gun} = 28^0$ (Fig. 3). This fit shows that when TRACE 3-D is used with a normalized emittance value of either 1 μm or 13 μm the beam envelope is consistent with the measured spot sizes, when the error bars are included.

Fig. 6. The normalized r.m.s. emittance for the 5 different values of the rf gun phase: (a) horizontal and (b) vertical. Measured emittance shown as vertical bars. PARMELA predicted emittances ($\varepsilon X = \varepsilon Y$) shown as diamonds. Note that the solenoids were set to a constant value (Table 1) during the phase sweep and thus the emittances are not, in general, compensated.

Fig. 7. PARMELA output values for the emittance ($\varepsilon$), the beam envelope ($\sigma$), and the space charge ratio divided by 100 ($R/100$) along the entire length of the AWA GTS beamline, from the photocathode at z = 0 to slightly beyond YAG-3. In the interval between YAG-1 and YAG-3 we see the beam at $\phi = 28^0$ is less space charge dominated than the one at $\phi = 48^0$.

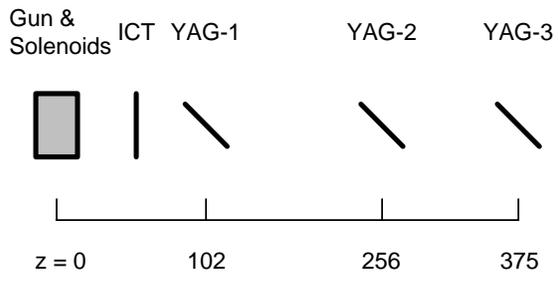

FIGURE 1

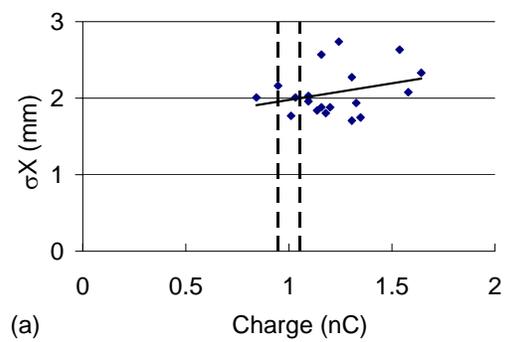

(a)

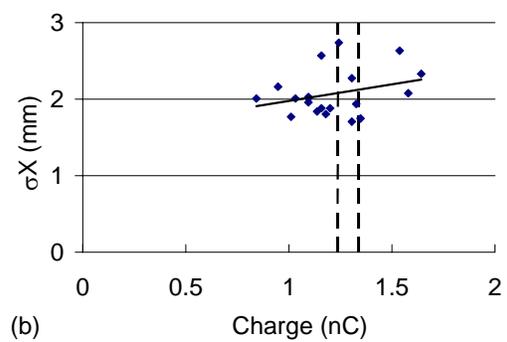

(b)

FIGURE 2

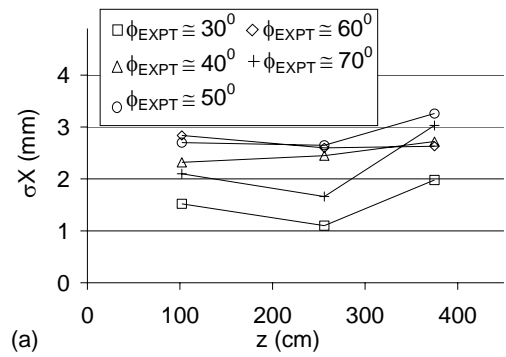

(a)

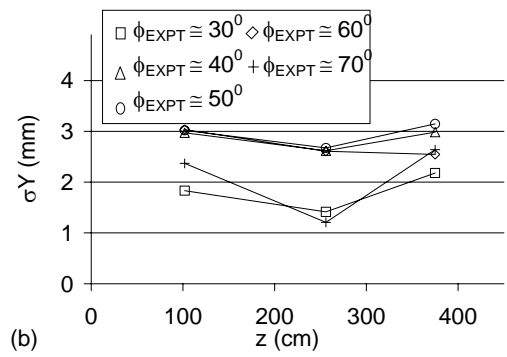

(b)

FIGURE 3

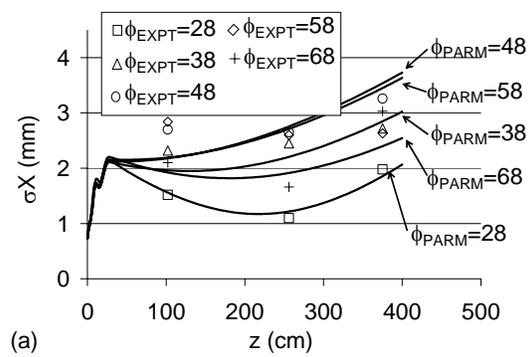

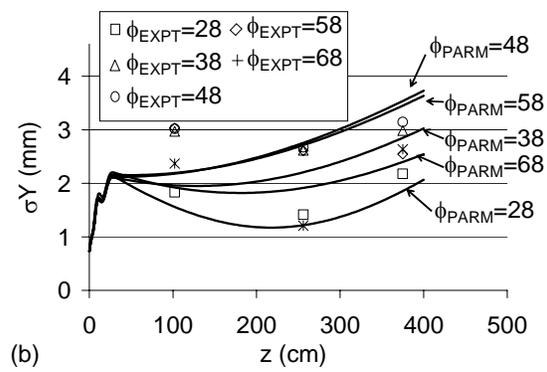

FIGURE 4

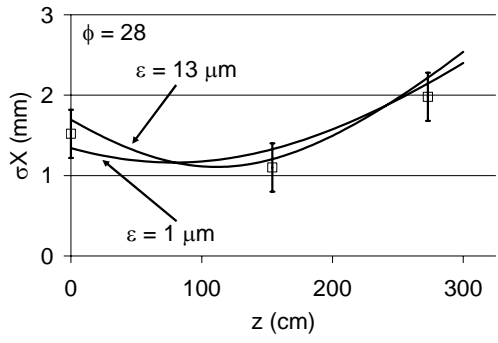

FIGURE 5

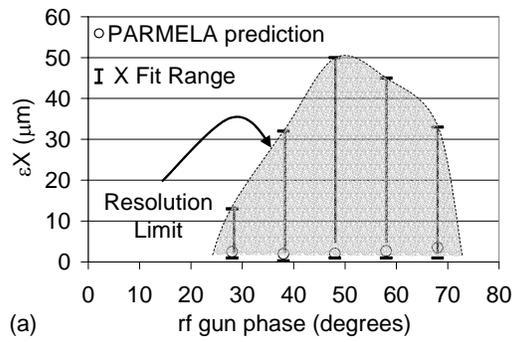

(a)

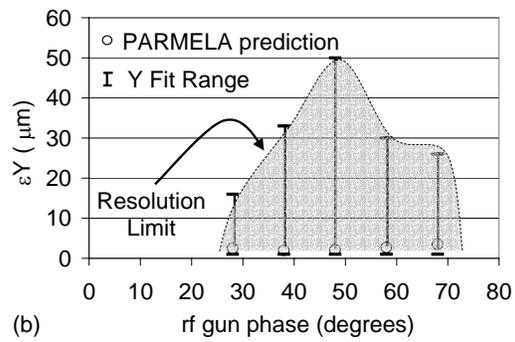

(b)

FIGURE 6

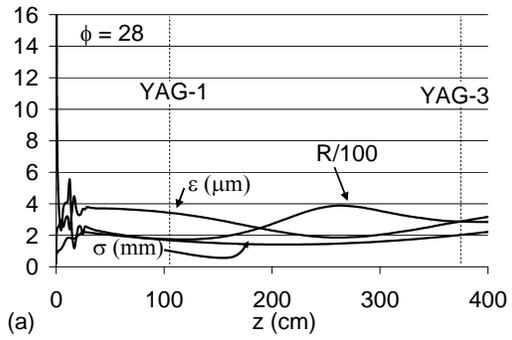

(a)

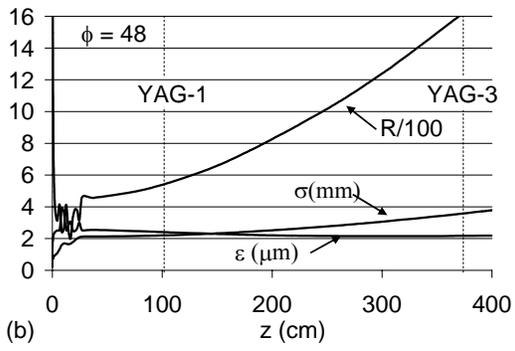

(b)

FIGURE 7

**Table 1**

Machine and beam parameters used during the measurements. $Q$ is the photoelectron charge; $R$ is the radius of the laser spot on the cathode; $B$, $F$, and $M$ are the Bucking, Focusing, and Matching solenoid field strengths on axis respectively; $E_{z0}$ is the peak field on-axis at the cathode; and $\phi_{gun}$ is the phase of the rf gun at launch.

| Parameter | Value |
|---|---|
| Nominal $Q$ (nC) | 1 |
| $R$ (mm) | 1.5 |
| $B$ (T) | 0.092 |
| $F$ (T) | 0.092 |
| $M$ (T) | 0.45 |
| $E_{z0}$ (MV/m) | 77 |
| $\phi_{gun}$ ($^0$) | $\cong 30^0$ to $70^0$ |

TABLE 1

**Table 2**

Longitudinal Phase Space Parameters taken from PARMELA simulation results using the input parameters shown in Table 1. Energy spread and bunch length values are the averaged values in the interval between YAG-1 and YAG-3.

| $\phi_{gun}$ | Energy (MeV) | r.m.s. Energy Spread (%) | Bunch Length (mm) |
|---|---|---|---|
| 28 | 7.66 | 0.52 | 0.82 |
| 38 | 7.95 | 0.63 | 0.83 |
| 48 | 8.07 | 0.88 | 0.85 |
| 58 | 8.00 | 1.2 | 0.87 |
| 68 | 7.70 | 1.7 | 0.95 |

TABLE 2